\documentclass{article}
\usepackage{arxiv}
\usepackage[utf8]{inputenc} 
\usepackage[T1]{fontenc}    
\usepackage[colorlinks=true,urlcolor=blue]{hyperref}       
\usepackage{url}            
\usepackage{booktabs}       
\usepackage{amsfonts}       
\usepackage{nicefrac}       
\usepackage{microtype}      
\usepackage{graphicx}
\usepackage{float}
\newcommand{\argmax}{\arg\!\max}

\title{Large-scale, Language-agnostic Discourse Classification of Tweets During COVID-19}

\author{ \vspace{0.2cm}
  Oguzhan~Gencoglu\\ 
  Faculty of Medicine and Health Technology, Tampere University\\
  Tampere, Finland \\
  \texttt{\url{oguzhan.gencoglu@tuni.fi}}
} 

\begin{document}
\maketitle

\begin{abstract}
Quantifying the characteristics of public attention is an essential prerequisite for appropriate crisis management during severe events such as pandemics. For this purpose, we propose language-agnostic tweet representations to perform large-scale Twitter discourse classification with machine learning. Our analysis on more than 26 million COVID-19 tweets shows that large-scale surveillance of public discourse is feasible with computationally lightweight classifiers by out-of-the-box utilization of these representations.
\end{abstract}

\keywords{text classification \and sentence embeddings \and Twitter \and natural language processing \and deep learning \and health informatics}

\section{Introduction}

Coronavirus disease 2019 (COVID-19) was declared a pandemic by the World Health Organization on 11 March 2020~\cite{cucinotta2020declares}. Since first recorded case in Wuhan, China in late December 2019, 45.6 million people have been infected by COVID-19 and consequently, 1.2 million people have lost their lives globally as of 30 October 2020~\cite{dong2020interactive}. This constitutes 700 times more deaths than SARS and MERS combined~\cite{mahase2020coronavirus}. During such large-scale adverse events, monitoring information seeking behaviour of citizens, understanding general overall concerns, and identifying recurring discussion themes is crucial for risk communication and public policy making~\cite{jurgens2018effect,van2020using}. This need is further amplified in a global pandemic such as COVID-19 as the primary responsibility of risk management is not centralized to a single institution, but distributed across society. For instance, a recent study by Zhong et al. shows that people's adherence to COVID-19 control measures is affected by their knowledge and attitudes towards it~\cite{zhong2020knowledge}. Previous national and global adverse health events show that social media surveillance can be utilized successfully for systematic monitoring of public discussion due to its instantaneous global coverage~\cite{signorini2011use,ji2013monitoring,ji2015twitter,weeg2015using,mollema2015disease,jordan2019using}.

Twitter, due to its large user-base, has been the primary social media platform for seeking, acquiring, and sharing information during global adverse events, including the COVID-19 pandemic~\cite{rosenberg2020twitter}. Especially during the early stages of the global spread, millions of posts have been tweeted in a span of couple of weeks~\cite{chen2020covid,gao2020naist,ieee2020data,aguilar2020dataset,chen2020tracking}. Consequently, several studies proposed and utilized Twitter as a data source for extracting insights on public health as well as insights on public attention during the COVID-19 pandemic. Focus of these studies include nowcasting or forecasting of the disease, sentiment analysis, topic modeling, and quantifying misinformation/disinformation. Due to the novelty and unknown epidemiological characteristics of COVID-19, accurate quantification of public discussions on social media becomes especially relevant for disaster management (e.g. devising timely interventions or clarifying common misconceptions).

So far, manual or automatic topical analyses of discussions on Twitter during COVID-19 pandemic have been performed in an \textit{exploratory} or \textit{descriptive} manner~\cite{abd2020top,rao2020retweets,park2020conversations}. Characterizing public discourse in these studies rely predominantly on manual inspection, aggregate statistics of keyword counts, or unsupervised topic modeling by utilizing joint distributions of word co-occurrences followed by qualitative assessment of discovered topics. Main reasons for previous studies to avoid supervised machine learning approaches can be lack of annotated (labeled) datasets of public discourse on COVID-19. Furthermore, previous studies either restrict their scopes to a single language (typically English tweets) or examine tweets from different languages in separate analyses. This is mainly due to limitations of traditional topic modeling algorithms as they usually do not operate in a multilingual or cross-lingual fashion. 

In this study, we propose large-scale characterization of public discourse themes by categorizing more than 26 million tweets in a supervised manner, i.e., classifying text into semantic categories with machine learning. For this purpose, we utilize two different annotated datasets of COVID-19 related questions and comments for training our algorithms. To be able to capture themes from 109 languages in a single model, we employ state-of-the-art multilingual sentence embeddings for representing the tweets, i.e., Language-agnostic BERT Sentence Embeddings (LaBSE)~\cite{feng2020language}. Our results show that large-scale surveillance of COVID-19 related public discourse themes and topics is feasible with computationally lightweight classifiers by out-of-the-box utilization of these representations. We release the full source code of our study along with the instructions to access the experiment datasets\footnote{\url{https://github.com/ogencoglu/Language-agnostic_BERT_COVID19_Twitter}}. We believe our work contributes to the pursuit of expanding social media research for disaster informatics regarding health response activities.

\section{Relevant Work}
\subsection{COVID-19 Twitter}

Content analysis of Twitter data has been performed by various studies during the COVID-19 pandemic. Some studies approach their research problem by manual or descriptive (e.g. n-gram statistics) content analysis of Twitter chatter for gaining relevant insights~\cite{park2020conversations,dewhurst2020divergent,thelwall2020retweeting,alshaabi2020world,hamamsy2020twitter,singh2020first,lopez2020understanding,kouzy2020coronavirus}, while other studies utilize unsupervised computational approaches such as topic modeling~\cite{abd2020top,rao2020retweets,wicke2020framing,jarynowski2020trends,ordun2020exploratory,medford2020infodemic,chen2020eyes,cinelli2020covid,hosseini2020content,jang2020exploratory,saad2020towards,odlum2020application,park2020risk,xue2020twitter,gupta2020covid,wang2020public,feng2020working,yin2020detecting,mcquillan2020cultural,omoya2020suspicion,sharma2020coronavirus,kabir2020coronavis}. A high percentage of studies performing topic modeling and topic discovery on Twitter utilize the well-established Latent Dirichlet Allocation (LDA) algorithm~\cite{rao2020retweets,wicke2020framing,medford2020infodemic,chen2020eyes,hosseini2020content,jang2020exploratory,park2020risk,xue2020twitter,gupta2020covid,wang2020public,feng2020working,yin2020detecting,mcquillan2020cultural,kabir2020coronavis}. Similar unsupervised approaches of word/n-gram clustering~\cite{saad2020towards,odlum2020application,omoya2020suspicion} or clustering of character/word embeddings~\cite{cinelli2020covid,sharma2020coronavirus} have been proposed as well.

Tweet data utilized for most of these studies are restricted to a single language. Majority of the studies restrict their analysis only to English tweets~\cite{abd2020top,thelwall2020retweeting,kouzy2020coronavirus,wicke2020framing,medford2020infodemic,jang2020exploratory,odlum2020application,xue2020twitter,wang2020public,feng2020working,yin2020detecting,mcquillan2020cultural,sharma2020coronavirus}, possibly exacerbating the already existing selection bias. Other studies have restricted their datasets to Japanese~\cite{omoya2020suspicion}, Korean~\cite{park2020conversations}, Persian/Farsi~\cite{hosseini2020content}, and Polish~\cite{jarynowski2020trends} tweets. While studies that collect multilingual tweets exist, they have conducted their analyses (e.g. topic modeling) separately for each language~\cite{dewhurst2020divergent,alshaabi2020world,park2020risk}.

\subsection{Representing Tweets}

As effective representation learning of generic textual data has been studied extensively in natural language processing research, tasks involving social media text benefit from recent advancements as well. While traditional feature extraction methods relying on word occurrence counts (e.g. \textit{bag-of-words} or \textit{term frequency-inverse document frequency}) have been extensively utilized in previous studies involving Twitter~\cite{rosa2011topical,kaleel2015cluster,lo2017unsupervised}, they have been replaced by distributed representations of words in a vector space (e.g. \textit{word2vec}~\cite{le2014distributed} or \textit{GloVe}~\cite{pennington2014glove} embeddings). Distributed word representations are learned from large corpora by a neural network, resulting in words with similar meanings being mapped to closer vector representations with a feature number that is much smaller than the vocabulary size. Consequently, sentences, documents, or tweets can be represented, e.g. as an average-pooling of its word embeddings. Such representations have also been learned specifically from Twitter corpora as \textit{tweet2vec}~\cite{vosoughi2016tweet2vec,dhingra2016tweet2vec} or \textit{hashtag2vec}~\cite{liu2018hashtag2vec}.

While distributed word/sentence embeddings provide effective capturing of semantics, they operate as a static mapping from the textual space to the latent space. Serving essentially as a dictionary look-up, they often fail to capture the context of the textual inputs (e.g. polysemy). This drawback has been circumvented by contextual word/token embeddings such as \textit{ELMo}~\cite{peters2018deep} or \textit{BERT}~\cite{devlin2019bert}. Contextual word embeddings enable the possibility of same word being represented as different vectors if it appears in different contexts. Several studies involving tweets utilized these deep neural network techniques or their variants either as a pre-training for further downstream tasks (e.g. classification, clustering, entity recognition) or for learning tweet representations from scratch~\cite{gencoglu2018deep,zhu2019iu,ray2019keyphrase,chowdhury2020identifying,roitero2020twitter,mazoyer2020french,nguyen2020bertweet,muller2020covid}. Even though BERT word embeddings are powerful as pre-trained language models for task-specific fine-tuning, Reimers et al. show that out-of-the-box sentence embeddings of BERT and its variants can not capture semantic similarities between sentences, requiring further training for that purpose~\cite{reimers2019sentence}. They propose a mechanism for learning contextual sentence embeddings using BERT neural architecture, i.e. \textit{sentence-BERT}, enabling large-scale semantic similarity comparison, clustering, and information retrieval with out-of-the-box vector representations~\cite{reimers2019sentence}. Studies involving Twitter data have been utilizing these contextual sentence embeddings successfully as well~\cite{gencoglu2020causal,baly2020written,kim2020leveraging,gencoglu2020cyberbullying}.

\section{Methods}
\subsection{Data}

\begin{figure}
\centerline{\includegraphics[width=\textwidth,trim={0 0 0 0},clip]{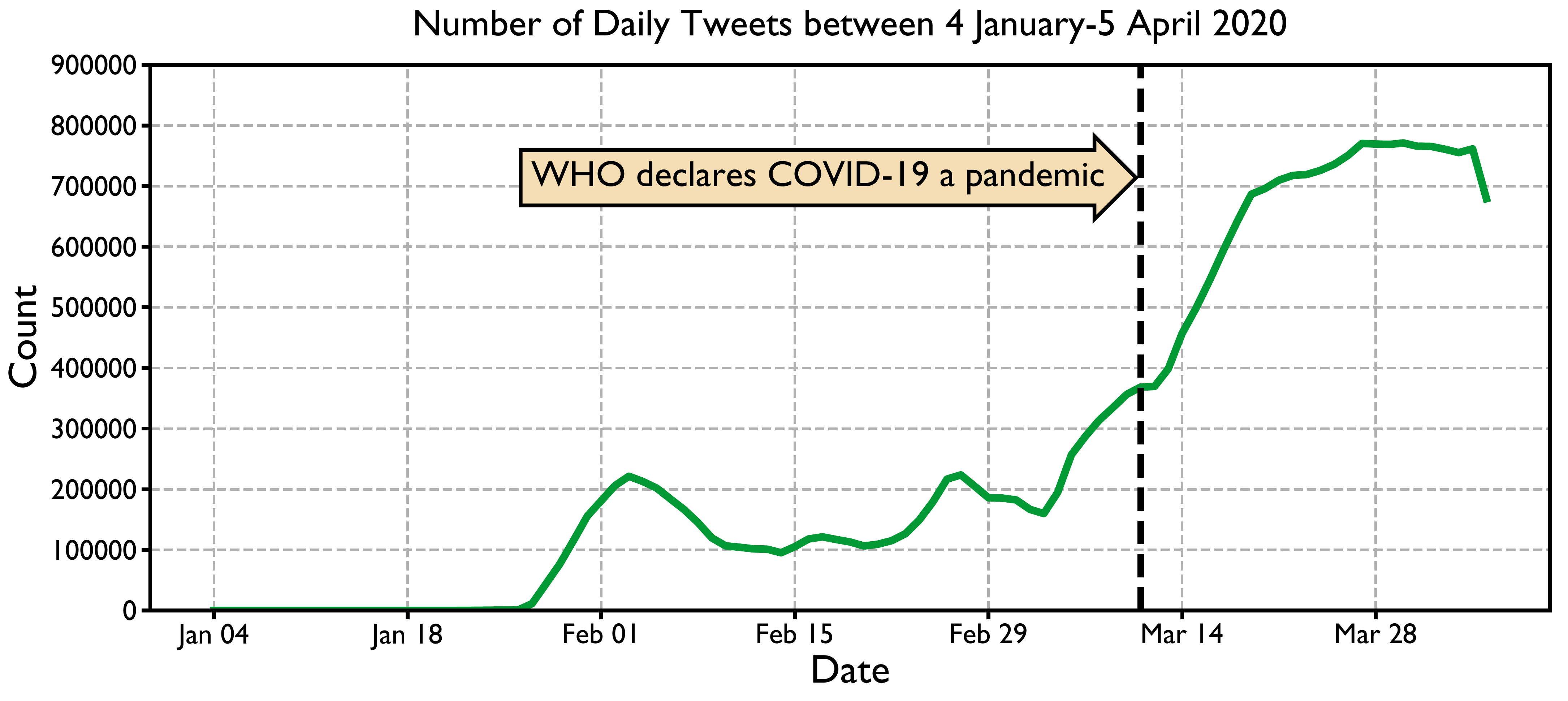}}
\caption{Daily Twitter activity related to COVID-19 during the early stages of the pandemic.}
\label{fig1}
\centering
\end{figure}

For Twitter data, we utilize the publicly available dataset of 152,920,832 tweets (including retweets) related to COVID-19 between the dates 4 January 2020 - 5 April 2020~\cite{banda2020twitter}. Tweets have been collected using the Twitter streaming API with the following keywords: \textit{COVID19}, \textit{CoronavirusPandemic}, \textit{COVID-19}, \textit{2019nCoV}, \textit{CoronaOutbreak}, \textit{coronavirus}, \textit{WuhanVirus}, \textit{covid19}, \textit{coronaviruspandemic}, \textit{covid-19}, \textit{2019ncov}, \textit{coronaoutbreak}, \textit{wuhanvirus}~\cite{panacealab}. As Twitter Terms of Service does not allow redistribution of tweet contents, only tweet IDs are publicly available. Extraction of textual content of tweets, timestamps, and other meta-data was performed with the use of open-source software Hydrator\footnote{\url{https://github.com/DocNow/hydrator}} with a Twitter developer account. For our study, we discard the retweets and at the time of extraction 26,759,164 unique tweets were available which is the final number of observations used in this study. Daily distribution of these tweets (7-day rolling average) can be observed from Figure \ref{fig1}.

For training machine learning classifiers, we utilize the following two recently-curated datasets: \textit{COVID-19 Intent}~\cite{arora2020cross} and \textit{COVID-19 Questions}~\cite{wei2020people}. Intent dataset consists of 4,938 COVID-19 specific utterances (typically a question or a request) categorized into 16 categories to describe the author's intent~\cite{arora2020cross}. For instance, the sample \textit{``is coughing a sign of the virus''} has an intent related to \texttt{Symptoms}. The dataset consists of English, French, and Spanish utterances and has been synthetically created by native-speaker annotators based on an ontology. We discard the uninformative categories of \texttt{Hi} and \texttt{Okay/Thanks} to end up with 4,325 samples from this dataset. We combine \texttt{Can\_i\_get\_from\_feces\_animal\_pets}, \texttt{Can\_i\_get\_from\_packages\_surfaces}, and \texttt{How\_does\_corona\_spread} categories into a single category of \texttt{Transmission}. Similarly, we merge \texttt{What\_if\_i\_visited\_high\_risk\_area} category into \texttt{Travel} category to end up with 11 categories (classes).

Questions dataset consists of 1,245 questions categorized into 16 categories collected from 13 sources~\cite{wei2020people}. 7 of the sources are frequently asked questions (FAQ) websites of recognized organizations such as the Center for Disease Control (CDC) and 6 of them are crowd-based sources such as Google Search. We use 594 samples from this dataset belonging to \texttt{Prevention}, \texttt{Reporting}, \texttt{Speculation}, \texttt{Symptoms}, \texttt{Transmission}, and \texttt{Treatment} categories. In the end, the dataset for our experiments, i.e., training and validating text classification algorithms, consists of 4,919 textual samples collected from the abovementioned two datasets. 11 category labels of the final dataset are \texttt{Donate}, \texttt{News \& Press}, \texttt{Prevention}, \texttt{Reporting},  \texttt{Share}, \texttt{Speculation}, \texttt{Symptoms}, \texttt{Transmission}, \texttt{Travel}, \texttt{Treatment}, \texttt{What Is Corona?}. Sample distribution of languages and categories among the dataset can be examined from Table \ref{table1} and Table \ref{table2}, respectively.

\begin{table}
\parbox{.45\linewidth}{
\centering
\begin{tabular}{cc}
\toprule
\textbf{Language} & \textbf{Samples} \\
\midrule
English & 2,119 \\
French & 1,400 \\
Spanish & 1,400 \\
\midrule
Total & 4,919 \\
\bottomrule
\end{tabular}
\caption{\label{table1} Distribution of languages. }
}
\hfill
\parbox{.45\linewidth}{
\centering
\begin{tabular}{cc}
\toprule
\textbf{Category} & \textbf{Samples} \\
\midrule
\texttt{Donate} & 310 \\
\texttt{News \& Press} & 310 \\
\texttt{Prevention} & 431 \\
\texttt{Reporting} & 389 \\
\texttt{Share} & 310 \\
\texttt{Speculation} & 363 \\
\texttt{Symptoms} & 348 \\
\texttt{Transmission} & 1,152 \\
\texttt{Travel} & 615 \\
\texttt{Treatment} & 381 \\ 
\texttt{What Is Corona?} & 310 \\
\midrule
Total & 4,919 \\
\bottomrule
\end{tabular}
\caption{\label{table2} Distribution of category labels. }
}
\end{table}

\subsection{Tweet Embeddings}

As the daily volume of COVID-19 related discussions on Twitter is enormous, computational public attention surveillance would benefit from lightweight approaches that can still maintain a high predictive power. Preferably, numerical representations should encode the semantics of tweets in such a way that simple vector operations should suffice for large-scale retrieval or even classification. Moreover, developed machine learning systems should be able to accommodate tweets in several languages to be able to capture the public discourse in an unbiased manner. Multilingual BERT-like contextual word/token embeddings~\cite{devlin2019bert} have been shown to be effective as pre-trained models if followed by a task-specific fine-tuning. However, they do not intrinsically produce effective sentence-level representations~\cite{reimers2019sentence}. In order to be able to take advantage of multilingual BERT encoders for extracting out-of-the-box sentence embeddings, we employ Language-agnostic BERT Sentence Embeddings~\cite{feng2020language}.

LaBSE embeddings combine BERT-based dual-encoder framework with masked language modeling (an unsupervised fill-in-the-blank task where a model tries to predict a masked word) to reach state-of-the-art performance in embedding sentences across 109 languages~\cite{feng2020language}. Trained on a corpus of 6 billion translation pairs, LaBSE embeddings provide out-of-the-box comparison ability of sentences even by a simple \textit{dot product} (essentially corresponding to \textit{cosine} similarity as embeddings are $l_2$ normalized). We encode both the training data and 26.8 million tweets using this deep learning approach, ending up with vectors of length 768 for each observation. Embeddings are extracted with \textit{TensorFlow} (version 2.2) framework in Python 3.7 on a 64 bit Linux machine with an NVIDIA Titan Xp GPU.

\subsection{Intent Classification}

As our choice of embeddings provide effective, out-of-the-box latent space representations of the textual data, simpler classifiers can be directly employed for identifying the prevalent topic of a tweet. In fact, LaBSE embeddings provide representations that are suitable to be compared with simple \textit{cosine similarity}~\cite{feng2020language}. We train 3 classifiers (multi-class, single-label classification), namely k-nearest neighbour (kNN), logistic regression (LR), and support vector machine (SVM) to classify the observations into 11 categories. We employ a 10-fold cross-validation scheme to evaluate the performance of the three models. For comparison, we run the same experiments for multilingual BERT embeddings (base, uncased) as well. Hyperparameters of the classifiers are selected by Bayesian optimization (see Section \ref{section3.4}). Once the embedding-classifier pair with its set of hyperparameters giving the highest cross-validation classification performance is selected, the classifier is trained with full dataset of 4,919 observations. With this model, inference on 26,759,164 samples of Twitter data embeddings is performed.

\subsection{Bayesian Hyperparameter Optimization}
\label{section3.4}

Typically, machine learning algorithms have several hyperparameters that require tuning for the specific task to avoid sub-optimal predictive performance. Most influential hyperparameters of k-nearest neighbour classifier are $k$ (number of neighbours) and distance metric (e.g. \textit{cosine}\footnote{Although, not an official distance metric as it violates triangle inequality.}, \textit{euclidean}, \textit{manhattan}, etc.). For support vector machine classifier, trade-off between training error and margin (essentially regularization), $C$, and the choice of kernel function (linear, polynomial, or radial basis function) are the most crucial hyperparameters. $l_2$ regularization coefficient is the main hyperparameter for logistic regression classifier. We formulate the problem of finding the optimal set of classifier hyperparameters, $\hat{\theta}$, as a Bayesian optimization problem:
\begin{equation}
\hat{\theta} = \argmax_{\theta} f(\theta),
\label{eq1}
\end{equation}
where $f(\theta)$ is the average of cross-validation accuracies for a given set of hyperparameters, i.e., $\frac{1}{N} \sum_{i=1}^{N} {ACC}_{i}$. For our experiments $N=10$ as we perform 10-fold cross-validation. We use \textit{Gaussian Processes} for the surrogate model~\cite{rasmussen2003gaussian} of the Bayesian optimization by which we emulate the statistical relationships between the hyperparameters and model performance, given a dataset. We run the optimization scheme for 30 iterations (each iteration corresponds to one full cross-validation) for each classifier separately.

Bayesian optimization is especially beneficial in settings where the function to be minimized/maximized, $f(\theta)$, is a black-box function without a known closed-form and expensive to evaluate~\cite{movckus1975bayesian}. As $f(\theta)$ corresponds to cross-validation performance in our case, it indeed is a black-box function that is computationally expensive to evaluate. That is our motive for employing Bayesian hyperparameter optimization instead of manual tuning or performing grid-search over a manually selected hyperparameter space.

\begin{figure}
\centerline{\includegraphics[width=0.7\textwidth,trim={0 0 0 0},clip]{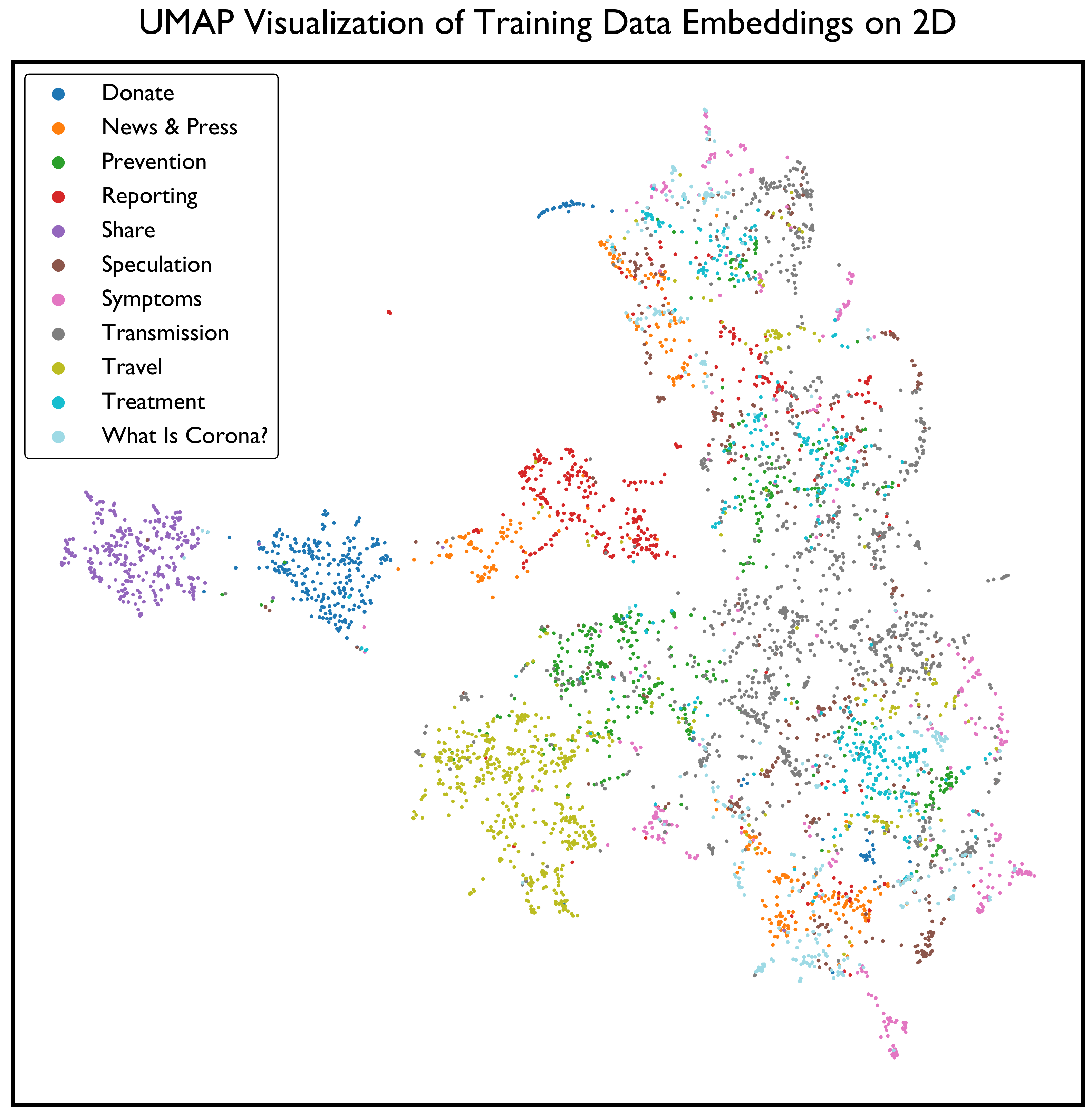}}
\caption{UMAP visualization of language-agnostic BERT sentence embeddings belonging to 4,919 observations among 11 COVID-19 discourse categories.}
\label{fig2}
\centering
\end{figure}

\subsection{Evaluation}

For visual inspection of embeddings, we utilize Uniform Manifold Approximation and Projection (UMAP) to map the high dimensional embeddings (768 for LaBSE and 1024 for BERT) to a 2-dimensional plane~\cite{mcinnes2018umap}. UMAP is a frequently used dimensionality reduction and visualization technique that can preserve global structure of the data better than other similar methods~\cite{mcinnes2018umap}. In their recent study, Ordun et al. employ UMAP visualization of COVID-19 tweets as well~\cite{ordun2020exploratory}.

Evaluation of classifiers and their sets of hyperparameters are performed by 10-fold cross-validation. Randomness (seed) in cross-validation splits are fixed in order to perform fair comparison. Average accuracy (\%) and F1 scores (micro and macro averages) across 10 folds are reported for all classifiers (for their best performing set of hyperparameters) for BERT and LaBSE embeddings. Confusion matrix for the best performing representation-classifier pair is reported as well. After running inference on Twitter data to classify 26.8 million tweets into 11 categories with the best performing classifier, we aggregate the overall distribution of Twitter chatter into percentages. We also show tweet examples from each predicted category. 

\section{Results}

UMAP visualization of LaBSE embeddings of the training data is depicted in Figure \ref{fig2}. Most visibly distinctive clusters belong to categories \texttt{Donate}, \texttt{Share}, and \texttt{Travel}. In this study, a cumulative of 88 hours of GPU computation was performed for extracting language-agnostic embeddings for the 26.8 million tweets which roughly corresponds to a carbon footprint of 9.5 kgCO$_2$eq (estimate by following~\cite{lacoste2019quantifying}).

\begin{table}
\centering
\begin{tabular}{c|ccc|ccc}
\toprule
& & \textbf{BERT} & & & \textbf{LaBSE} & \\ 
\cmidrule{2-7}
\textbf{Model} & \textbf{Accuracy (\%)} & \textbf{F1 (micro)} & \textbf{F1 (macro)} & \textbf{Accuracy (\%)} & \textbf{F1 (micro)} & \textbf{F1 (macro)}\\
\midrule
kNN & 72.54 & 0.725 & 0.725 & 82.76 & 0.828 & 0.827\\
LR  & 76.62 & 0.766 & 0.771 & 86.05 & 0.844 & 0.846\\
SVM & 81.81 & 0.818 & 0.820 & \textbf{86.92} & \textbf{0.876} & \textbf{0.881}\\
\bottomrule
\end{tabular}
\caption{\label{table3} Cross-validation results of three classifiers for BERT and LaBSE embeddings. }
\end{table}

\begin{figure}
\centerline{\includegraphics[width=0.8\textwidth,trim={0 0 0 0},clip]{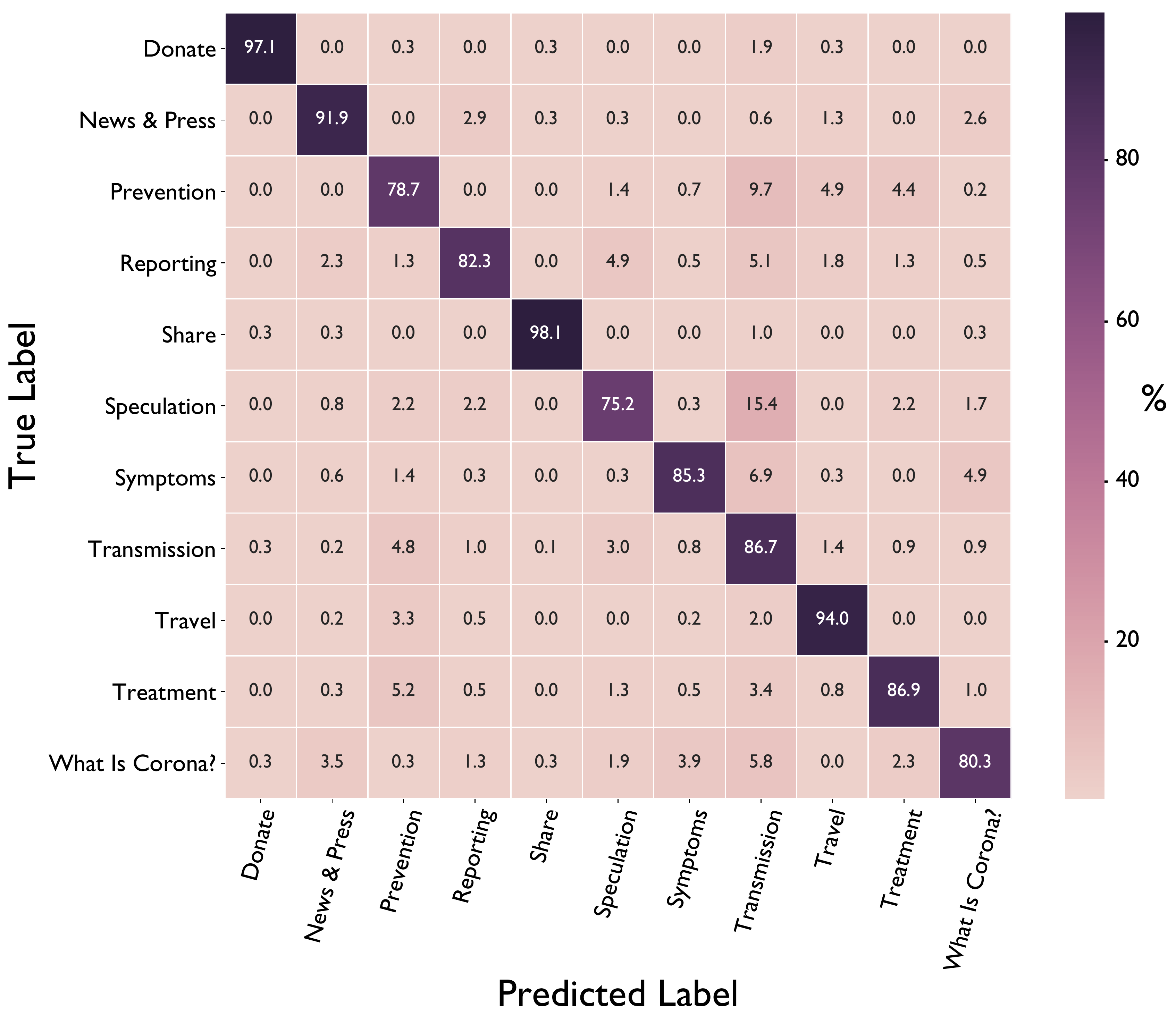}}
\caption{Normalized confusion matrix of SVM classifier predictions across cross-validation folds.}
\label{fig3}
\centering
\end{figure}

10-fold cross-validation results for the classifiers with the highest scoring set of hyperparameters are shown in Table \ref{table3}. For all classifiers, LaBSE embeddings outperform multilingual BERT embeddings. Best hyperparameters for k-nearest neighbour classifier were found to be $k=7$ and \textit{cosine} distance for both embeddings. Optimal regularization coefficients for logistic regression was found to be $4.94 \times 10^{3}$ for LaBSE and $1.01 \times 10^{2}$ for BERT representations. For support vector machine, optimal $C$ was found to be 5.07 and 3.51 for LaBSE and BERT representations, respectively. Best choice of kernel function was found to be radial basis function for both representations. Best performing classifier was found to be support vector machine classifier applied on LaBSE embeddings with 86.92~\% accuracy, 0.876 micro-F1 score, and 0.881 macro-F1 score. Confusion matrix of this classifier out of cross-validation predictions can be examined from Figure \ref{fig3}. In parallel to visual findings on Figure \ref{fig2}, \texttt{Donate}, \texttt{Share}, and \texttt{Travel} classes reach high accuracies of 97.1~\%, 98.1~\%, and 94.0~\%, respectively. Classifier has the highest error rate for the \texttt{Prevention} and \texttt{Speculation} classes, both staying below 80~\% accuracy. Our results show that more than 15~\% of samples belonging to \texttt{Speculation} category have been misclassified as \texttt{Transmission}.

\begin{figure}
\centerline{\includegraphics[width=\textwidth,trim={0 0 0 0},clip]{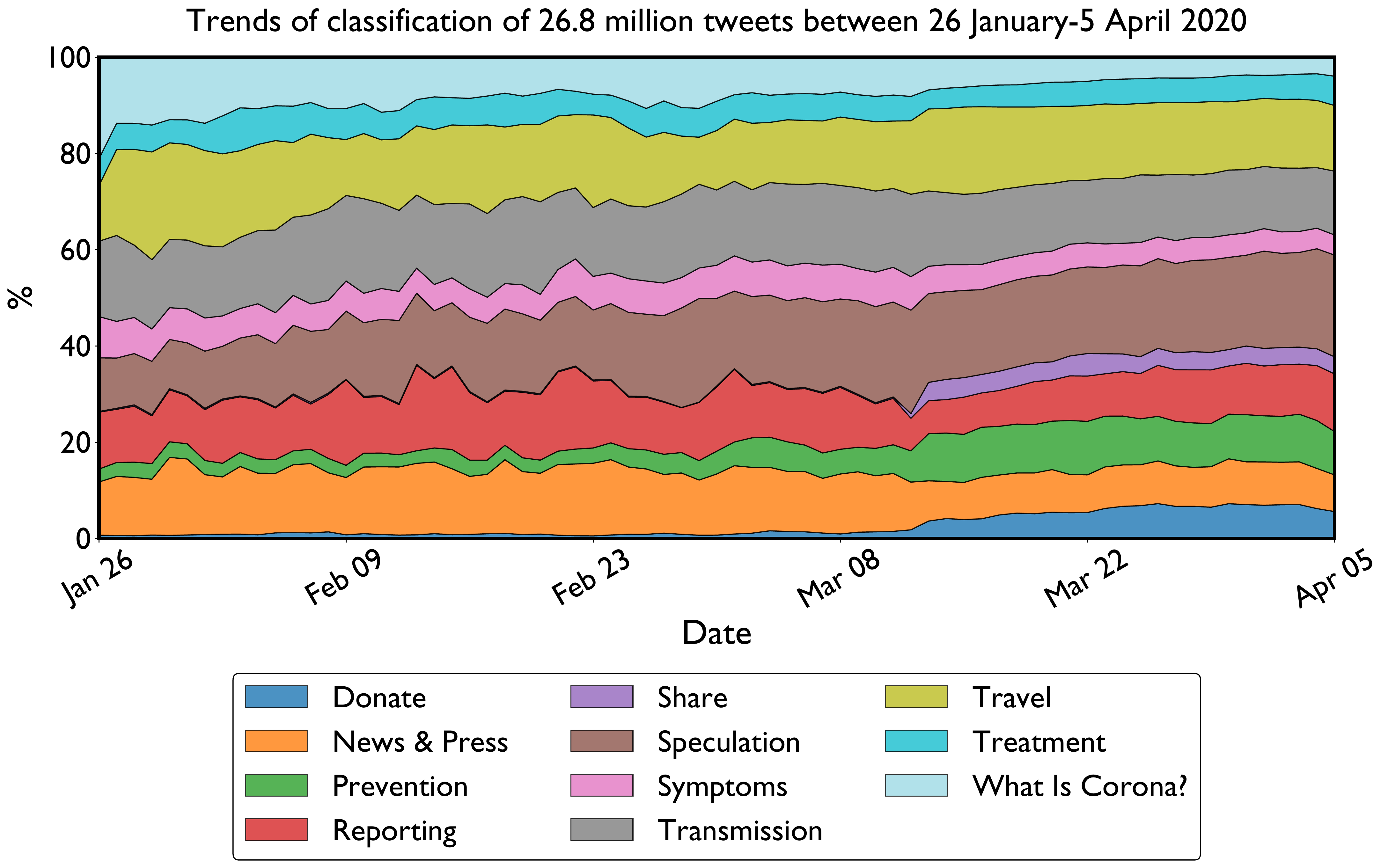}}
\caption{Distribution of semantic discussion categories in Twitter predicted by the classifier during COVID-19.}
\label{fig4}
\centering
\end{figure}

Figure \ref{fig4} depicts the timeline of normalized daily category distributions obtained by running inference on tweets posted between 26 January and 5 April 2020. Transmission and travel-related chatter as well as speculations (opinions on origin of COVID-19, myths, and conspiracies) show significance presence throughout the pandemic. \texttt{What Is Corona?}, i.e. questions and inquiries regarding what exactly COVID-19 is, shows a presence in the early stages of the pandemic but decreases through time, possibly due to gained scientific knowledge about the nature of the disease. On the contrary, prevalence of \texttt{Prevention} related tweets increase through time especially after the declaration of pandemic by WHO on March 11. Similarly, chatter for \texttt{Donation} discussions are observed only starting from March. Timeline curves become smoother (less spiky) with increasing date as the percentage changes between consecutive days gets smaller. This is intuitive as the total number of tweets in January is several magnitudes lower than that of April and sudden percentage jumps in January can be attributed to only a handful of tweets. Finally, random samples of tweets and their predicted labels can be observed from Table \ref{table4}.

\begin{table}
\centering
\begin{tabular}{p{8.5cm}p{2.8cm}}
\toprule
\textbf{\centerline{Tweet}}  & \textbf{\centerline{Predicted Class}} \\
\midrule
\normalsize China Providing Assistance To Pakistani Students Trapped in Wuhan: Ambassador - \#Pakistan & \normalsize \texttt{Donate} \\
\midrule
\normalsize Results are in. State health officials say three suspected cases of Coronavirus have tested NEGATIVE. There is a forth possible case from Washtenaw County being sent to the CDC. & \normalsize \texttt{News \& Press} \\
\midrule
\normalsize what are good steps to protect ourselves from the Coronavirus? & \normalsize \texttt{Prevention} \\
\midrule
\normalsize The first coronavirus case has been confirmed in the U.S. \#virus & \normalsize \texttt{Reporting}  \\
\midrule
\normalsize Share this and save lives \#coronavirus \#SSOT & \normalsize \texttt{Share}  \\
\midrule
\normalsize \#coronavirus Don’t let these ignorant people make you believe that this corona virus is any different than SARS IN 2003 which was contained after a few months. They want you to panic as they have ulterior motives such as shorting the stock market etc. & \normalsize \texttt{Speculation}  \\
\midrule
\normalsize  I have a rushing sound in my ears. It doesn't seem to match the symptoms for the \#coronavirus so perhaps it is the sound of the \#EU leaving my body... & \normalsize \texttt{Symptoms} \\
\midrule
\normalsize what animals can carry Wuhan coronavirus? & \normalsize \texttt{Transmission} \\
\midrule
\normalsize can we ban flights from wuhan pls?!? & \normalsize \texttt{Travel}  \\
\midrule
\normalsize ¿Qué medicamento nos colará en está ocasión la industria farmacéutica para combatir al coronavirus? & \normalsize \texttt{Treatment} \\
\midrule
\normalsize Oque é coronavirus? & \normalsize \texttt{What Is Corona?} \\
\bottomrule
\end{tabular}
\caption{\label{table4} Example tweets and predicted classification categories. }
\end{table}

\section{Discussion}

Adequate risk management in crisis situations has to take into account not only the threat itself but also the perception of the threat by the public~\cite{sandman1993responding}. In digital era, public heavily relies on social media to inform their level of risk perception, often in a rapid manner. In fact, social media enhances collaborative problem-solving and citizens’ ability to make sense of the situation during disasters~\cite{jurgens2018effect}. With this paradigm in mind, we attempt to perform large-scale classification of 26.8 million COVID-19 tweets using natural language processing and machine learning. We utilize state-of-the-art language-agnostic tweet representations coupled with simple, lightweight classifiers to be able to capture COVID-19 related discourse during a span of 13 weeks.

Our first observation of ``increasing Twitter activity with increased COVID-19 spread throughout the globe'' (Figure \ref{fig1}) is in parallel with other studies. For instance, Bento et al. show that Internet searches for ``coronavirus'' increase on the day immediately after the first case announcement for a location~\cite{bento2020evidence}. Wong et al. correlates announcement of new infections and Twitter activity~\cite{wong2020paradox}. Similar associations have been discovered between official cases and Twitter activity by causal modeling as well~\cite{gencoglu2020causal}. Secondly, we show that language-agnostic embeddings can be utilized in an out-of-the-box fashion, i.e., without requiring task-specific fine-tuning of BERT models. A SVM classifier reaches 86.92~\% accuracy and 0.881 macro-F1 score for classification into 11 topic categories. Finally, we show that overall public discourse shifts through the pandemic. Questions of ``what coronavirus is'' leave their place to donation and prevention related discussions as the disease spreads into more and more countries especially during March 2020. Tweets related to donation increase especially around 13 March 2020 when WHO and the United Nations Foundation start a global COVID-19 donation fund~\cite{responsefund}.

When compared to existing studies that often employ unsupervised topic modeling, our approach tries to perform public attention surveillance with a more automated perspective as we formulate the problem as a supervised learning one. Topic modeling with LDA, which has been employed by majority of previous studies, relies on manual and qualitative inspection of discovered topics. Furthermore, plain LDA fails to accommodate contextual representations and does not assume a distance metric between discovered topics as it is based on the notion that words belonging to a topic are more likely to appear in the same document. With language-agnostic embeddings, we also include tweets from languages other than English to our analysis, hence decrease the selection bias.

Utilization of large-scale social media data for extracting health insights is even more pertinent during a global pandemic such as COVID-19, as running randomized control trials becomes less practical. Moreover, traditional surveys for public attention surveillance may further stress the participants whose mental health and overall well-being might have been affected by lockdowns, associated financial issues, and changes in social dynamics~\cite{wang2020immediate,cullen2020mental,brooks2020psychological}. Once accurate estimation of global or national discourse is possible, social media can also be used to direct people to trusted resources, counteract misinformation, disseminate reliable information, and enable a culture of preparedness~\cite{merchant2020social}. Assessment of effectiveness of public risk communication and interventions is also feasible with properly designed computational systems. Guided by machine learning insights, some of these interventions can be made on social media itself.

Our study has several limitations. First, the training data consists of single label annotations while in reality a tweet can have several topics simultaneously, e.g. \texttt{Prevention} and \texttt{Travel}. Secondly, we do not employ a confidence threshold for categorizing tweets which forces our model to classify every observation into one of the 11 categories. Considering some Twitter discourses related to COVID-19 may not be properly represented by our existing categories, a probability threshold can be introduced for the final classification decision. Finally, we discard retweets in our analysis, which in fact contributes to public attention on Twitter.

Future research includes running similar analysis for a more granular category set or sub-categories. For instance, \texttt{Speculation} category can be divided into conspiracies related to origin of the disease, transmission characteristics, and treatment options. Including up-to-date Twitter data (after April 2020) as well as extracting location-specific insights will be performed in future analyses as well.

\section{Conclusions}

Transforming social media data into actionable knowledge for public health systems face several challenges such as advancing methodologies to extract relevant information for health services, creating dynamic knowledge bases that address disaster contexts, and expanding social media research to focus on health response activities~\cite{chan2019challenges}. We hope our study serves this purpose by proving methodologies for large-scale, language-agnostic discourse classification on Twitter.





\bibliographystyle{unsrt}
\bibliography{references}

\end{document}